\documentstyle[11pt]{article}
\def\fnote#1#2{\begingroup\def\thefootnote{#1}\footnote{#2}\addtocounter
{footnote}{-1}\endgroup}
\def\BM#1{\mbox{\boldmath{$#1$}}}

\begin{document}

\hfill{UTTG-20-18}

\vspace{20pt}

\begin{center}
{\large {\bf {Soft Bremsstrahlung}}}

\vspace{20pt}

Steven Weinberg\fnote{*}{Electronic address:
weinberg@physics.utexas.edu}\\
{\em Theory Group, Department of Physics, University of
Texas\\
Austin, TX, 78712}

\vspace{30pt}

\noindent
{\bf Abstract}
\end{center}

Simple analytic formulas are considered  for the energy radiated in low frequency bremsstrahlung from fully ionized gases.  A  formula that has been frequently cited over many years turns out  to have only a limited range of validity, more narrow than for a formula derived using the Born approximation.    In an attempt  to find a more widely valid simple formula,  a soft photon theorem is employed, which in this context implies that the differential rate of photon emission in an electron-ion collision with definite initial and final electron momenta is correctly given for sufficiently soft photons by the Born approximation, to all orders in the Coulomb potential.  Corrections to the Born approximation arise because the upper limit on photon energy for this theorem to apply to a given collision becomes increasingly stringent as the scattering approaches the forward direction.  A general formula is suggested that takes this into account.

\noindent

\vfill

\pagebreak

\begin{center}
{\bf I INTRODUCTION}
\end{center}

The emission of radio waves from hot ionized interstellar gas is largely due to soft bremsstrahlung, the radiation of a low energy photon in the deflection  of a free electron with much larger kinetic energy by the Coulomb field of an atomic nucleus.  It is conventional to express the  rate $j(\nu,v)$ of energy emission per time, per photon solid angle, and per photon frequency interval at frequency $\nu$  from an electron of velocity ${\bf v}$  with $|{\bf v}|=v$  
due to bremsstrahlung in a 
 fully ionized gas  as the approximate classical electrodynamics result given  in 1923 by Kramers[1], times a  ``free-free Gaunt factor'' $g_{\rm ff}(\nu,v)$ that incorporates quantum and  other corrections:
\begin{equation}
j(\nu,v)\equiv \frac{8 \pi Z^2e^6n_{\cal I}}{3\sqrt{3}\,c^3 m_e^2 v}g_{\rm ff}(\nu,v)\;,
\end{equation}
where $n_{\cal I}$ is the number density of ions, $Ze$ is the ionic charge (with $e$ everywhere in unrationalized electrostatic units), and $m_e$ is the electron mass.  For an ionized gas in kinetic equilibrium at temperature $T$, this gives the emissivity, the rate of radiation energy emitted per time, per volume, per photon solid angle, and per photon frequency interval:
\begin{equation}
j_\nu(T)=\int_{m_e{\bf v}^2/2>h\nu} d^3v\;n_e({\bf v},T)\,j(\nu,|{\bf v}|)
\end{equation}
where $n_e({\bf v},T)d^3v$ is the number density of free electrons with velocity  in a range $d^3v$ at ${\bf v}$.  Using the Maxwell-Boltzmann distribution for  $n_e({\bf v},T)$, this gives the emissivity
\begin{equation}
j_\nu(T)=
\frac{8 Z^2e^6n_{\cal I}n_e}{3  c^3 (k_{\cal B}T)^{1/2}m_e^{3/2}}\left(\frac{2\pi}{3}\right)^{1/2}\overline{g}_{\rm ff}(\nu,T)\;.
\end{equation}
where  $n_e$ is the total number density of free electrons; $k_{\cal B}$ is the Boltzmann constant; and 
$\overline{g}_{\rm ff}(\nu,T)$ is the thermally averaged free-free Gaunt factor (briefly, the thermal Gaunt factor):
\begin{equation}
\overline{g}_{\rm ff}(\nu,T)=\frac{m_e}{k_{\cal B}T}\int_{\sqrt{2h\nu/m_e}}^\infty g_{\rm ff}(\nu,v)\;\exp\left(-\frac{m_ev^2}{2k_{\cal B}T}\right)\;v\,dv\;.
\end{equation}

Astrophysicists today chiefly rely on various numerical calculations (e.g., refs. [2], [3], [4], [5]) of the 
Gaunt factor, based on a set of  quite complicated formulas:
\begin{eqnarray}
&& g_{\rm ff}(\nu,v)=\frac{2\sqrt{3}}{\pi\xi\xi'}\Big[(\xi^2+\xi'^2+2\xi^2\xi'^2)I_0
-2\xi\xi'(1+\xi^2)^{1/2}(1+\xi'^2)^{1/2}I_1\Big]I_0\;,\nonumber\\&&{}
\end{eqnarray}
where
\begin{eqnarray*}
&&I_\ell=\frac{1}{4}\left(\frac{4\xi\xi'}{(\xi'-\xi)^2}\right)^{\ell+1}e^{\pi(\xi+\xi')/2}\frac{|\Gamma(\ell+1+i\xi)\Gamma(\ell+1+i\xi')|}{\Gamma(2\ell+1)}\nonumber\\&&
\times \left(\frac{\xi+\xi'}{\xi'-\xi}\right)^{-i\xi-i\xi'}{}_2F_1\Bigg(\ell+1-i\xi,\,\ell+1-i\xi';\,2\ell+2;\,-\frac{4\xi\xi'}{(\xi'-\xi)^2}\Bigg)\;.\nonumber\\{}
\end{eqnarray*}
Here $\xi\equiv Ze^2/\hbar v$ and  $\xi'\equiv Ze^2/\hbar v'$, with  $v'$  the magnitude of the final electron velocity, given in terms of $\nu$ and $v$ by the condition of energy conservation
\begin{equation}
m_ev'^2/2=m_ev^2/2-h\nu\;.
\end{equation}
Also, ${}_2F_1$ is a confluent hypergeometric function, with power series expansion
$$
{}_2F_1(a,b;c;x)=\sum_{n=0}^\infty \frac{(a)_n\,(b)_n}{(c)_n}\frac{x^n}{n!}\;,
$$
where for any complex $z$
$$ (z)_n\equiv z(z+1)\cdots (z+n-1)\;{\rm for}\; n=1,2,3\dots;~~~~(z)_0\equiv 1\;.$$
These are
 derived[2] from the  summation of a partial wave expansion[6] of results originally given by Sommerfeld[7]  Still, it would be  useful to have a widely valid simple  analytic formula for the  Gaunt factor, in order easily to see trends in how it varies with  various parameters, and in order easily to calculate the thermal Gaunt factor (4).  Above all,  from an independent derivation of a simple analytic formula we can gain a  more detailed physical understanding of what is going on in the bremsstrahlung process.  

A simple formula for the thermal Gaunt factor has been often given, without providing a derivation,  in treatises  on the interstellar medium (for example, [8], [9], [10], [11])).
\begin{equation}
 \overline{g}_{\rm ff}(\nu,T)=\frac{\sqrt{3}}{\pi}\left[\ln\left(\frac{(2k_{\cal B}T)^{3/2}}{\pi Z e^2 \nu m^{1/2}_e}\right)-\frac{5\gamma}{2}\right] \;,
\end{equation}
where $\gamma$ is the Euler constant, $\gamma=0.577\dots$.   Some of these references indicate that the formula holds for photons that are soft, in the sense that $h\nu\ll k_{\cal B}T$.   (They also note that it is necessary to assume that  the photon frequency is much larger than the plasma frequency $\nu_P$, so that Debye screening can be ignored.  This is not a stringent condition, and will be taken for granted throughout.)  But none suggest that there are more stringent conditions on the frequency and temperature for the formula to be a valid approximation.

For this formula Spitzer[8]  cited an article by  Scheuer[11], who found Eq.~(7) by a purely classical calculation of the emissivity per electron, which gave a result
\begin{equation}
g_{\rm ff}(\nu,v)=\frac{\sqrt{3}}{\pi}\left[\ln\left(\frac{m_ev^3}{\pi Z e^2\nu}\right)-\gamma\right]\;,
\end{equation}
As Scheuer found, using this in Eq.~(4) gives the widely quoted thermal Gaunt factor (7) for $h\nu\ll k_{\cal B}T$. .

It is  not  possible  that Eq.~(8) could be a good approximation for general photon frequencies and electron velocities with  $h\nu\ll m_ev^2/2$ .  
Contrary to Eq.~(8), if 
$\xi\equiv Ze^2/ \hbar v $  is much less than unity (so that the Coulomb potential at an electron-ion separation equal to the electron de Broglie wave length is much less than the electron kinetic energy)   one would expect  $j(\nu,v)$ to have a finite limit, given by the Born approximation --- that is, keeping in the matrix element only terms of first order in the Coulomb potential:
\begin{equation}
g^{\rm Born}_{\rm ff}(\nu,v)=\frac{\sqrt{3}}{\pi}\ln\left(\frac{2m_ev^2}{h\nu}\right)\;.
\end{equation}
(The derivation of Eq.~(9) for $\xi\ll 1$ is given in the next section.)  Since Eq.~(8) does not reduce to Eq.~(9) for $\xi\ll 1$, where Eq.~(9) applies,  it cannot be correct for small $\xi$.  It also cannot be correct when $\xi$ is much larger than the ratio of electron to photon energies, because there it gives a negative Gaunt factor.   Accordingly, the formula (7) for the thermal Gaunt factor derived from  Eq.~(8) cannot be expected to hold for general photon frequencies and electron temperatures with $h\nu\ll k_{\cal B}T$. 

Recently Albalat and Zimmerman[13] have shown that Eq.~(8) follows from the ``exact'' formula (5) used in numerical calculations, for $\xi$ in the range 
\begin{equation}
1 \ll \xi \ll m_ev^2/h\nu \;.
\end{equation}
(They subsequently found that, though it seems to have been largely forgotten, in 1962 a review article[14] had obtained the same result.)  
For instance, for  $2h\nu/m_ev^2=10^{-3}$, Eq.~(8) is within a few percent of  numerical results[5] for  $\xi$ between 1 and 10.  On the other hand, for $2h\nu/m_ev^2=10^{-2}$ the range in which (8) agrees with numerical results is vanishingly narrow.  

In contrast, the Born approximation (9) agrees very well with numerical calculations where $\xi< 1$, and  the Scheuer Gaunt factor (8) does not.
For instance, for $h\nu=10^{-3}m_ev^2/2$, Eq.~(9) gives $g^{\rm Born}_{\rm ff}(\nu,v)=4.573$, while numerical calculations[5] give $g_{\rm ff}(\nu,v)$ equal to 4.5730 for  $\xi=10^{-3}$ and for $\xi=10^{-2}$,  dropping only to 4.5672 for $\xi=0.1$ and to 4.2093 for $\xi=1$.  (An electron has $\xi<1$ if its kinetic energy is larger than the binding energy of a $1s$ atomic electron.)  In contrast, for the same ratio of photon and electron energies,  Eq.~(8) gives a Gaunt factor that is 76\% too large for $\xi=10^{-3}$, and still 21\% too large for  $\xi=0.1$.  

This leaves us with the task of finding an approximate analytic formula for the Gaunt factor that is generally valid for $\xi>1$.  Section II derives what I think is a new formula for the matrix element for soft bremsstrahlung, valid for any $\xi$ to all orders in the Coulomb potential.  This formula leads immediately to the Born approximation (9) in the case $\xi\ll 1$.  To deal with more general values of $\xi$, a general soft-photon  theorem  is used in Section III to show that the 
the differential rate of photon emission in an electron-ion collision with definite initial and final electron momenta is correctly given for sufficiently soft photons by the Born approximation, to all orders in the Coulomb potential.
Nevertheless, as explained in Section IV, integration over the final electron direction introduces corrections to the Born approximation for the Gaunt factor for $\xi>1$.  Properties of the general formula for the bremsstrahlung matrix element derived in Section II suggest a framework for a more general formula.

\begin{center}
{\bf II A GENERAL FORMULA}
\end{center}

To derive the Born approximation result (9) for $\xi\ll 1$, and to understand the decrease of the Gaunt factor below the Born approximation value for $\xi$ of order unity and greater,  it will be useful first to provide what I think is a new formula for the matrix element for bremsstrahlung that is valid to all orders in the Coulomb potential. 

Taking electrons to be non-relativistic, which also entails the electric dipole approximation for the interaction of electrons with the quantized electromagnetic field, the term in the matrix element (that is, the coefficient of the energy conservation delta function in the S-matrix)  for bremsstrahlung of first order in this interaction and to all orders in the Coulomb interaction between an electron and an ion is given in the ``distorted wave Born approximation''[15] as
\begin{equation}
M=\frac{-2\pi i}{\sqrt{2qc}(2\pi \hbar)^{3/2}}\times \frac{-\sqrt{4\pi}e\hbar^2}{m_e}\int d^3r\; \psi'^*({\bf r}){\bf e}^*(\hat{q},\lambda)\cdot\BM{\nabla}\psi({\bf r})
\end{equation}
where  ${\bf e}(\hat{q},\lambda)$ is the polarization vector (with ${\bf e}^*\cdot{\bf e}=1$) for a photon with momentum  ${\bf q}$  and helicity $\lambda$, and $\psi$ and $\psi'$ are respectively ``in'' and ``out'' normalized solutions of the Schr\"{o}dinger equation for the initial and final electrons.      If we multiply $M$ with $qc=m_ev^2/2-m_ev'^2/2$ and use the  Schr\"{o}dinger equations for $m_ev^2\psi/2$ and $m_ev'^2\psi'/2$, we find by an integration by parts that the kinetic energy terms in the Schr\"{o}dinger equations cancel, while   the potential terms  cancel except where the gradient in Eq.~(9) acts on the electron-ion interaction potential $V$, so 
\begin{equation}
M=\frac{-ie\sqrt{\hbar}}{(qc)^{3/2}m_e}\int d^3r\;\psi'^*({\bf r})\;{\bf e}^*(\hat{q},\lambda)\cdot \Big[\BM{\nabla}V({\bf r})\Big]\psi({\bf r})\;.
\end{equation}
Using the general rules for calculating rates in quantum mechanics, and setting $qc=h\nu$, the rate of emission of radiation energy per time, per photon solid angle, per photon frequency interval, and per final electron solid angle when an electron is scattered from initial velocity ${\bf v}$ to final velocity ${\bf v}'$ 
 is then 
\begin{equation}
j(\nu,{\bf v}\rightarrow  {\bf v}')=h\nu\times\frac{h^5\nu^2 n_{\cal  I}m_e^3}{4\pi c^3}\int d^2\hat{q}\, \int_0^\infty v'^2\,dv'\sum_\lambda |M|^2   \delta\left(\frac{m_ev^2}{2}-\frac{m_ev'^2}{2}-h\nu\right)\;.
\end{equation}

Eqs.~(12) and (13) apply for  a general potential $V$.
For a Coulomb potential $V(r)=-Ze^2/r$. the exact wave functions are well known:
\begin{eqnarray}
&&\psi({\bf r})=\frac{\Gamma(1+i\xi)e^{-\xi\pi/2}}{(2\pi \hbar)^{3/2}}e^{i{\bf k}\cdot {\bf r}}\;{}_1F_1\Big(-i\xi;1;(ik|{\bf r}|-i{\bf k}\cdot {\bf r})\Big)\;,\nonumber\\
&&\psi'({\bf r})=\frac{\Gamma(1-i\xi')e^{-\xi'\pi/2}}{(2\pi \hbar)^{3/2}}e^{i{\bf k}'\cdot {\bf r}}\;{}_1F_1\Big(i\xi';1;(-ik'|{\bf r}|-i{\bf k}'\cdot {\bf r})\Big)\;,
\end{eqnarray}
where ${\bf k}\equiv {\bf v}m_e/\hbar$ and ${\bf k}'\equiv {\bf v}'m_e/\hbar$; ${\bf v}$ and ${\bf v}'$ are the initial and final electron velocities;   $\xi\equiv Ze^2/\hbar v$; $\xi'\equiv Ze^2/\hbar v'$; and ${}_1F_1$ is another confluent hypergeometric function.

So far, this is valid for non-relativistic electrons  to all orders in the Coulomb potential $V(r)$.    If we now take $\xi\ll 1$ (so that for soft photons also $\xi'\ll 1$) then the functions ${}_1F_1$  in the wave functions (11) take the form $1+O(\xi)$ and $1+O(\xi')$\, and the matrix element (10) is then given by just the term of first order in $V$ [16]:
\begin{equation}
M=\frac{4\pi Ze^3\hbar^{3/2}}{(h\nu)^{3/2}(2\pi\hbar)^3 m^2_e}\frac{({\bf v}-{\bf v}')\cdot {\bf e}^*(\hat{q},\lambda)}{({\bf v}-{\bf v}')^2}
\end{equation} 
Using this in Eq.~(11), we find the emission rate per electron:
\begin{equation}
j^{\rm Born}(\nu,{\bf v}\rightarrow{\bf v}')=\frac{4Z^2e^6n_{\cal I}v'}{3\pi c^3m_e^2} \frac{1}{|{\bf v}-{\bf v}'|^2}
\;.
\end{equation}
(with $v'$ given by the energy conservation condition $m_ev'^2/2=m_ev^2/2-h\nu$).  
Integrating over the final electron direction gives, for soft photons,
\begin{eqnarray}
&& j^{\rm Born}(\nu,v)=\int d^2\hat{v}' j^{\rm Born}(\nu,{\bf v}\rightarrow{\bf v}')=\frac{8Z^2e^6n_{\cal I}}{3c^3m_e^2v}\ln\left(\frac{v+v'}{v-v'}\right)\nonumber\\&&~~~~~~\rightarrow \frac{8Z^2e^6n_{\cal I}}{3c^3m_e^2v}\ln\left(\frac{2m_ev^2}{h\nu}\right)
\end{eqnarray}
corresponding to  the Gaunt factor (9), in disagreement with Scheuer's formula (8).  It has recently been shown [12] that Eq.~(9) also follows  in the limit $\xi\rightarrow 0$  from the formula (5) used in refs.~[2]-[5]. 	.

Using Eq.~(9) in Eq.~(4)  gives a thermal Gaunt factor
\begin{equation}
\overline{g}_{\rm ff}(\nu,T)=\frac{\sqrt{3}}{\pi}\Bigg[\ln\Big(4k_{\cal B}T/h\nu\Big)-\gamma\Bigg]\;,
\end{equation}
which we expect to be valid if   $2h\nu/ m_e v_T^2\ll 1$ and $Ze^2/ \hbar v_T\ll 1$, where $v_T\equiv \sqrt{2k_{\cal B}T/m_e}$ is a typical thermal velocity.  
There is a possible problem in this thermal averaging: no matter how small $2h\nu/ m_e v_T^2$ and $Ze^2/ \hbar v_T$ may be, there will always be some  electrons with $v\ll v_T$ for which  $2h\nu/ m_e v^2$ and/or $Ze^2/ \hbar v$ are not small.  But the effect of these slow electrons is suppressed by $d^3v/v=4\pi v\,dv$ in the velocity integration (4).  In any case,  Scheuer's Eq.~(8) already disagrees with the Born approximation result (9) before thermal averaging, under circumstances in which the Born approximation is valid.

\begin{center}
{\bf III THE SOFT PHOTON THEOREM}
\end{center}

We need to go further, and understand the changes in the Gaunt factor when $\xi\equiv Ze^2/\hbar v$ is of order unity or greater.
At first sight, it might seem that the Born approximation should continue to apply for soft photons whatever the value of $\xi$.  This is because the emission per electron $j(\nu,{\bf v}\rightarrow{\bf v}')$ for fixed electron directions is correctly given in the soft photon limit $v'\rightarrow v$ by the Born approximation result (16),   to all orders in the Coulomb potential, whether or not $\xi$ is small.  
      
This conclusion is based on a very general low energy theorem[17] of quantum electrodynamics.  According to this theorem, which is valid to all orders in perturbation theory, the differential rate for a general process $\alpha\rightarrow \beta$ with emission of any number of soft photons with total energy less than some amount $E$  is
given in the soft photon limit $E\rightarrow 0$ by 
\begin{equation}
d\Gamma_{\alpha\rightarrow \beta}(<E)\rightarrow (E/\Lambda)^A\,b(A)\,d\Gamma^0_{\alpha\rightarrow \beta}\;.
\end{equation}
Here
\begin{equation}
A=-\frac{1}{8\pi^2\hbar c}\sum_{n,m}\frac{4\pi\eta_n\eta_m e_ne_m}{\beta_{nm}}\ln\left(\frac{1+\beta_{nm}}{1-
\beta_{nm}}\right)\;,
\end{equation}
where the sums run over all particles participating in the reaction $\alpha\rightarrow\beta$;  $e_n$ is the charge of the $n$th particle; $\eta_n$ equals   $+1$ or $-1$ for  particles in the initial state $\alpha$ or final state $\beta$; and $c\beta_{nm}$  is the  velocity of either of particles $n$ or $m$ in the rest  frame of the other particle:
\begin{equation}
\beta_{nm}\equiv \left[1-\frac{m_n^2m_m^2c^4}{(p_n\cdot p_m)^2}\right]^{1/2}\;.
\end{equation}
Also, $b(A)$ is the function
\begin{equation}
b(A)\equiv \frac{1}{\pi}\int_{-\infty}^{+\infty}  \frac{\sin \sigma\,d\sigma}{\sigma}\exp\left[A\int_0^1\frac{d\omega}{\omega}\Big(e^{i\omega\sigma}-1\Big)\right]=1-\frac{\pi^2A^2}{12}+\dots\;,
\end{equation}
and 
$ d\Gamma^0_{\alpha\rightarrow \beta} $ is the differential rate for the same process without soft photon emission and without radiative corrections from virtual infrared photons, where $\Lambda$ is a more-or-less arbitrary upper limit on virtual photon four-momenta that is used to define what we mean by ``infrared.''  (As we shall see, $\Lambda$ will not appear in the non-relativistic limit relevant to this paper.)  The differential rates $d\Gamma_{\alpha\rightarrow \beta}(<E)$ and $d\Gamma^0_{\alpha\rightarrow \beta}$ are rates for producing the particles in the final state $\beta$ in some infinitesimal element of their momentum spaces, the same for both rates.  (The formula given in ref. [17] has been modified here by inserting a factor $4\pi$ in Eq.~(20) to account for the use here of unrationalized units for electric charge, and inserting a factor $1/\hbar c$ to make $A$ dimensionless in cgs units.)

This is no place to re-derive this old result, but it may be useful here to remark that it applies because the insertion of a soft-photon external line of momentum ${\bf q}$ in any external line for a charged particle in the process $\alpha\rightarrow\beta$ produces an internal line connecting this vertex to the rest of the diagram; the propagator for this line contributes a $1/q$ singularity for $q\rightarrow 0$, whose residue is  proportional to the matrix element for the process without the soft photon.  This accounts for a factor $1/q$ in Eq.~(12), which multiplies the kinematic factor $1/\sqrt{q}$ already present in Eq.~(11).  (For photon absorption, the corresponding $1/q^{3/2}$ factor in the matrix element accounts for a factor $(k_{\cal B}T)^{-3}$ in the Kramers opacity for free-free transitions.)  These diagrams dominate the matrix element for $q\rightarrow 0$, because insertion of the soft photon line in an internal line of the process  $\alpha\rightarrow\beta$ does not produce this pole.  

Formula (19) for the soft photon emission rate applies for relativistic or non-relativistic processes involving particles of arbitrary spin, whatever the interactions may be that produce the reaction $\alpha\rightarrow\beta$.  It is considerably simplified if we specialize to the non-relativistic case, for which in some reference frame all velocities of the particles in the states $\alpha$ and $\beta$ are much less than $c$.  In this case all $\beta_{nm}$ are much less than one, and we can use the expansion
\begin{equation}
\frac{1}{\beta_{nm}}\ln\left(\frac{1+\beta_{nm}}{1-
\beta_{nm}}\right)=2+\frac{2}{3}\beta_{nm}^2+\dots\;.
\end{equation}
The first term does not contribute in Eq.~(20), because the conservation of electric charge gives $\sum_n \eta_ne_n=0$.  Hence in the non-relativistic case, Eq,~(20) becomes
\begin{equation}
A= -\frac{1}{3\pi\hbar c}\sum_{n,m}\eta_n\eta_m e_ne_m \beta^2_{nm}\;,
\end{equation}
This is at  most of order $v^2/c^2$, so in the non-relativistic limit $b(A)=1$, and Eq.~(19) becomes
\begin{equation}
d\Gamma_{\alpha\rightarrow \beta}(<E)\rightarrow \left[1+A\ln\left(\frac{E}{\Lambda}\right)\right]\,d\Gamma^0_{\alpha\rightarrow \beta}\;.
\end{equation}
The rate of energy radiation per frequency interval  is then simply
\begin{equation}
h\nu\frac{d}{d\nu}d\Gamma_{\alpha\rightarrow \beta}(<h\nu)\rightarrow hA\,d\Gamma^0_{\alpha\rightarrow \beta}\;.
\end{equation}

If we now specialize to the case of a non-relativistic electron of velocity ${\bf v}$ scattered by a Coulomb potential into a  final velocity ${\bf v}'$, the sum in Eq.~(26) will be dominated by terms in which particles $n$ and $m$ are respectively the initial and final electron, or vice versa.  This gives
\begin{equation}
A=\frac{2e^2}{3\pi \hbar c^3}\Big|{\bf v}-{\bf v}'|^2\;.
\end{equation}
If needed we could use this in Eq.~(24) with any assumption about the differential rate $d\Gamma^0_{{\bf v}\rightarrow {\bf v}'}$ for electron scattering without photon emission, taking account of complications like screening  or finite ion size in scattering by atoms, but the calculation of $d\Gamma^0_{{\bf v}\rightarrow {\bf v}'}$ beyond low orders of perturbation theory would then   be complicated.   Fortunately, as well known,  if we limit ourselves to the scattering of electrons by the Coulomb field of an unscreened heavy point ion of charge $Ze$, then the differential Coulomb scattering rate per electron $d\Gamma^0_{{\bf v}\rightarrow {\bf v}'}$ is correctly given to all orders in the Coulomb potential by the Born approximation result:
\begin{equation}
d\Gamma^0_{{\bf v}\rightarrow {\bf v}'}=\frac{4Z^2e^4 n_{\cal I} v}{m_e^2|{\bf v}-{\bf v}'|^4}d^2\hat{v}'\;,
\end{equation}
where  $v\equiv |{\bf v}|$, and $d^2\hat{v}'$ is the solid angle into which the electron is scattered.
Using Eqs.~(27) and (28) in Eq.~(26), the differential rate of energy radiation in soft bremsstrahlung per photon frequency interval, per photon solid angle,  and per electron is
\begin{equation}
j(\nu,{\bf v}\rightarrow {\bf v}')d^2\hat{v}'\equiv \frac{h\nu}{4\pi}\frac{d}{d\nu}d\Gamma_{{\bf v}\rightarrow {\bf v}'}(<h\nu)\rightarrow \frac{4Z^2e^6n_{\cal I}v}{3\pi c^3 m_e^2\Big|{\bf v}-{\bf v}'\Big|^2}d^2\hat{v}'\;,
\end{equation}
just as in the Born approximation result (14).  Unfortunately, as we shall see, although the soft photon theorem tells us that Eq.~(29) holds for any $\xi$ and  any fixed initial and final electron velocities and sufficiently small photon frequency, for $\xi>1$ the upper bound  on the photon frequency for its  validity becomes increasingly stringent as the final electron direction approaches the initial direction.

\begin{center}
{\bf IV DEPARTURES FROM THE BORN APPROXIMATION}
\end{center}

Departures from the  Born approximation for soft photons arise because, to calculate the emission per electron  we need to integrate over the 
 final electron direction.  In the strict soft photon limit, in which $\nu=0$ and $v=v'$, there is a logarithmic divergence in the integral, arising from the configuration in which $\hat{v}'$ is parallel to $\hat{v}$.  In the Born approximation, for $\nu$ small but non-zero the integral is cut off by the inequality of $v$ and $v'$, yielding Eq.~(17), from which the   Gaunt factor (9) follows as before.  But  beyond the Born approximation, does Eq.~(15) correctly describe the behavior of the emission rate when $v'\neq v$, where the soft photon theorem does not apply?

To answer this, we note 
 that the singularity in  the photon emission rate  when ${\bf v}'\rightarrow {\bf v}$ arises entirely from the  slow decrease of the Coulomb potential at large $r$.  To evaluate this singularity, we use the well-known asymptotic behavior of the Coulomb wave functions: for $r\rightarrow \infty$
\begin{equation}
\psi({\bf r})\rightarrow (2\pi\hbar)^{-3/2}e^{i{\bf k}\cdot {\bf r}}  \; |kr-{\bf k}\cdot {\bf r}|^{i\xi}\;,~~~
\psi'({\bf r})\rightarrow (2\pi\hbar)^{-3/2}e^{i{\bf k}'\cdot {\bf r}}  \; |k'r+{\bf k}'\cdot {\bf r}|^{-i\xi'}\;,
\end{equation}
where ${\bf k}={\bf v}m_e/\hbar$ and ${\bf k}'={\bf v}'m_e/\hbar$.
 Using these asymptotic forms in Eq.~(12), we see that when $ |v'-v|/v\simeq h\nu/m_ev^2$ and the angle $\theta$ between ${\bf v}$ and ${\bf v}'$  are both very small, the singularity in the matrix element, is of the form
\begin{equation}
M\rightarrow   \frac{-iZe^3\sqrt{h}}{(h\nu)^{3/2}m_e}\frac{(m_e/\hbar)^{i\xi+i\xi'}}{(2\pi\hbar)^3}\int d^3r \left(\frac{{\bf e}^*\cdot {\bf r} }{r^3}\right)\exp\Big(i{\bf r}\cdot ({\bf v}-{\bf v}')(m_e/\hbar)\Big)|vr-{\bf v}\cdot {\bf r}|^{i\xi}|v'r+{\bf v}'\cdot {\bf r}|^{i\xi'}\;.
\end{equation}
This can be straightforwardly calculated to leading order in $ h\nu/m_ev^2$ and $\theta$ in two limiting cases:

First, if $ h\nu/m_ev^2\ll \theta\ll 1$ we encounter a singularity:
\begin{equation}
M\rightarrow \frac{4\pi Ze^3\hbar^{3/2}}{(h\nu)^{3/2}(2\pi\hbar)^3 m^2_e}\frac{v^{2i\xi}({\bf v}-{\bf v}')\cdot {\bf e}^*}{|{\bf v}-{\bf v}'|^{2+2i\xi}}
\times \frac{\Gamma(1+2i\xi)\Gamma\Big(\frac{1}{2}-i\xi\Big)\cosh\xi\pi}{\sqrt{\pi}\Gamma\Big(1-i\xi\Big)}\;.
\end{equation}
This is not the same as in the Born approximation matrix element (15), which is no surprise, because  the Coulomb scattering amplitude itself is  affected by higher orders in the  Coulomb potential.  But here as in Coulomb scattering the higher-order corrections in Eq.~(32) are just phases, which do not appear in $|M|^2$, so the singularity in  the integrand of the emission per electron is of the form
\begin{equation}
|M|^2\rightarrow |M_{\rm Born}|^2\;.
\end{equation}
The soft photon theorem tells us that this is also true for $ h\nu/m_ev^2\ll 1$  even where  the angle $\theta$ between electron directions is not small and the integral (10) is not dominated by large $r$.

Second, moving away from the case covered by the soft photon theorem, if $  \theta \ll h\nu/m_ev^2 \ll 1$,  we find a singularity
\begin{equation}
M\rightarrow \frac{4\pi Ze^3\hbar^{3/2}}{(h\nu)^{3/2}(2\pi\hbar)^3 m_e^2}\frac{v^{2i\xi}( {\bf v}- {\bf v'})\cdot {\bf e}^*}{
|{\bf v}-{\bf v}'|^{2+2i\xi}}
 \Gamma(1+2i\xi)\Gamma\Big(\frac{1}{2}-i\xi\Big)\Gamma\Big(1+i\xi\Big)\frac{\cosh\xi\pi}{\sqrt{\pi}}\;.
\end{equation}
Here the corrections to the Born approximation are not just phases.  Instead,
\begin{equation}
 |M|^2\rightarrow |M_{\rm Born}|^2\times \frac{\pi^2\xi^2}{\sinh^2\pi\xi}\;.
\end{equation}

For a rough approximation to the Gaunt factor when $\xi$ is not small, we introduce a critical angle $\theta_c$, and tentatively suppose that Eq.~(33) holds for $\theta>\theta_c$ and 
Eq.~(35) holds for $\theta<\theta_c$.   Then the emission rate per electron is
\begin{equation}
j(\nu,v)=\frac{4Z^2e^6n_{\cal I}v'}{3\pi c^3 m_e^2}\left[\int_{\theta>\theta_c} \frac{d^2\hat{v}'}{\Big|{\bf v}-{\bf v}'\Big|^2}+\frac{\pi^2\xi^2}{\sinh^2\pi\xi}\int_{\theta<\theta_c} \frac{d^2\hat{v}'}{\Big|{\bf v}-{\bf v}'\Big|^2}\right]\;,
\end{equation}
corresponding to a Gaunt factor 
\begin{equation}
g_{\rm ff}(\nu,v)=\frac{\sqrt{3}}{\pi}\left[\ln\left(\frac{2m_ev^2}{h\nu\zeta}\right)
+\frac{\pi^2\xi^2}{\sinh^2\pi\xi}\ln\zeta\right]\;,
\end{equation}
where
\begin{equation}
\zeta\equiv (1+2(\theta^2_c/(2h\nu/m_ev^2)^2)^{1/2}\;.
\end{equation}
For $\xi\ll 1$ the factor $\pi^2\xi^2/\sinh^2\pi\xi$ is close to unity, the dependence of the Gaunt factor on the unknown function $\zeta$ drops out, and we recover the Born approximation (9).  For  $\xi>1$ the factor $\pi^2\xi^2/\sinh^2\pi\xi$  is exponentially small, and this Gaunt factor reduces to the form (9) of the Born approximation, except for the factor $1/\zeta$ in the argument of the logarithm.  Since $\zeta>1$, the Gaunt factor for $\xi>1$ is always less than the Born approximation value.  In the limited range (10) where Eq.~(8) is valid, we have $\zeta\simeq \xi e^\gamma\gg 1$, and so here the critical angle is $\theta_c\simeq  \xi e^\gamma \sqrt{2}h\nu/m_ev^2$. 
  More generally, the decrease in the Gaunt factor found in numerical calculations for $\xi>1$ is evidently due to a depletion of photon radiation from nearly forward electron scattering.

I am grateful to Paul Shapiro for helpful conversations about bremsstrahlung in astrophysics, and to Aaron Zimmerman and Sergi Albalat for informative discussions of numerical calculations.  This article is based on work supported by the National Science Foundation under Grant Number PHY-1620610, and with support from the Robert A. Welch Foundation, Grant No. F-0014.

\begin{center}
REFERENCES
\end{center}
\begin{enumerate}
  \item H. Kramers, Phil. Mag. {\bf 46}, 836 (1923).
  \item W. J. Karzas and R. Latter, Astrophys. J. Suppl. {\bf 6}, 167 (1961)
  \item  D. G. Hummer, Astrophys. J. {\bf 327}, 472 (1988).
  \item  R. S. Sutherland, Mon. Not. Roy. Ast. Soc. {\bf 300}, 321 (1998).
  \item  P. A. M. van Hoof {\em et al.},  Mon. Not. Roy. Ast. Soc. {\bf 444}, 420 (2014).
  \item  L. C. Biedenharn, Phys. Rev. {\bf 102}, 262 (1955).
  \item A. J. Sommerfeld, {\em Atombau und Spektralinien}, Vol. II, 
Chapter 7, Section 5. (Vieweg \& Sohn, Braunschweig, 1939).
\item D. E. Osterbrock,   {\em Astrophysics of Gaseous Nebulae and Active Galactic Nuclei} (University Science Books, Mill Valley, CA, 1989).
\item L. Spitzer, Jr.  {\em Physical Processes in the Interstellar Medium} (John Wiley \& Sons, Inc., New York, 1998).  
  \item B. T. Draine, {\em Physics of the Interstellar and Intergalactic Medium} (Princeton University Press, Princeton, NJ, 2011).
  \item    W. J. Maciel,   {\em Astrophysics of the Interstellar Medium}, transl. M. Serote Roos (Springer Sciences, New York, 2013).  
  \item F. A. G. Scheuer,    Mon. Not. Roy. Astron. Soc. {\bf 120}, 231 (1960).  This formula was cited as a result of classical theory by  L. Oster,  Rev. Mod. Phys. {\bf 33}, 525 (1961).
  \item  S. N. Albalat and A. Zimmerman, paper in preparation.
  \item   P. J. Brussaard  and H. C. van de Hulst,   Rev. Mod. Phys. {\bf 34}, 507 (1962). 
  \item For textbook discussions of the distorted wave Born approximation, photon emission interactions, and Coulomb wave functions, see for instance S. Weinberg, {\em Lectures on Quantum Mechanics}, 2nd ed. (Cambridge University Press, Cambridge, UK, 2015), sections 8.6, 11.7, and 7.9.
  \item Essentially the same Born approximation matrix element was obtained by a direct use of old-fashioned second order perturbation theory in a calculation of the inverse bremsstrahlung contribution to opacity by H-Y. Chiu, {\em Stellar Physics} (Blaisdell Publishing Co., Waltham, MA, 1968)), Sec. 5.11.
\item S. Weinberg,   Phys. Rev.  {\bf 140}, B516 (1965). 
  
\end{enumerate}

\end{document}